\documentclass[12pt]{article}
\usepackage{amsmath}
\usepackage{amsfonts}
\usepackage{amssymb}
\usepackage{latexsym}
\usepackage{graphicx}
\usepackage[english]{babel}
\input epsf.sty

\begin{document}

\title{
{\normalsize \hskip4.2in USTC-ICTS-07-14} \\{A Stochastic Measure
for Eternal Inflation}}

\vspace{3mm} \author{{Miao Li$^{1,2}$\footnote{mli@itp.ac.cn}, Yi Wang$^{2,1}$\footnote{wangyi@itp.ac.cn}}\\
{\small $^{1}$ The Interdisciplinary Center for Theoretical Study}\\
{\small of China (USTC), Hefei, Anhui 230027, P.R.China}\\ {\small
$^{2}$ Institute of Theoretical Physics, Academia Sinica, Beijing
100080, P.R.China}} \date{} \maketitle
\begin{abstract}
We use the stochastic approach to investigate the measure for slow
roll eternal inflation. The probability for the universe of a given
Hubble radius can be calculated in this framework. In a solvable model, it is
shown that the probability for the universe to evolve
from a state with a smaller Hubble radius to that of a larger Hubble
radius is dominated by the classical probability without the
stochastic source. While the probability for the universe to evolve
from a larger Hubble radius to a smaller one is suppressed by
$\exp(-\Delta S)$, where the de Sitter entropy $S$ arises naturally
in this stochastic approach.
\end{abstract}

\newpage

\section{Introduction}
The inflation paradigm has proven to be remarkably successful
in solving the problems in the standard hot big bang cosmology
\cite{Guth81,Linde82,Steinhardt82,Starobinsky-inf}. Inflation also
predicts that fluctuations of quantum origin were generated
and frozen to seed wrinkles in the cosmic microwave background (CMB)
\cite{CMBobserve,WMAP} and today's large scale structure
\cite{Mukhanov81,Guth82,Hawking82,Starobinsky82,Bardeen83}.

In a usual inflation model, if the universe starts at a high energy
scale, inflation should be eternal to the future
\cite{steinhardt-nuffield,vilenkin-eternal,linde-eternal}. There are
two classes of eternal inflation models. One of them is
characterized by the slow-rolling nature. During the eternal stage
of inflation, the amplitude of quantum fluctuation of the inflaton
field is comparable to its classical motion. Such large fluctuations
make the universe fall into self-reproduction process and prevent
the energy density from decreasing. So inflation will never end
globally. Another class of eternal inflation models is characterized
by forming bubbles of one vacuum within another. Once the decay rate
of the false vacuum is smaller than the Hubble scale, the spatial
volume of the false vacuum is increasing faster than the decay of
the false vacuum volume. Then inflation becomes eternal to the
future.

It is widely believed that eternal inflation is indeed happening in
the universe, and we just live in a local reheated domain of the
eternal inflating universe. So it is important to study eternal
inflation precisely and try to make predictions from the eternal
inflation scenario.

Unfortunately, it is rather difficult to describe eternal inflation
precisely. There are several open problems in the attempts to
describing eternal inflation, for example, the measure
problem and the initial condition problem.

%and the problem of the ability for prediction.

The key problem of eternal inflation is how to construct a measure
for the eternal inflation \cite{vilenkin-measure, Bousso:2006ev,
Gibbons:2006pa, Linde:2006nw, Podolsky:2007vg, Li:2007rp,
Cai:2007et}. One of the difficulties is how to
construct such a measure preserving symmetry of general
relativity, and staying finite despite the fact that several kinds
of infinities frequently occur in a naive construction. To overcome
this difficulty, people have proposed two kinds of ansatz, namely, the
``global'' measure \cite{vilenkin-measure, Linde:2006nw} and the
``local'' measure \cite{Bousso:2006ev}.

In the global approach, infinities are regularized by imposing
cutoffs, neverthless some cutoff-independent results can be obtained.
The global measure encompasses the physics
separated by event horizons, so it contradicts the holographic principle in
a fundamental way \cite{Goheer:2002vf}. To counter this, a local measure
describing the physics seen by a comoving
observer was proposed. This approach is based on the cosmic complementarity principle
and as a bonus, it does not suffer from infinities. The main proponent of this
approach is Bousso \cite{Bousso:2006ev}. Bousso and collaborators played their
game with models whose dynamics is governed by tunneling processes, they have not
studied a concrete model with a definite Lagrangian.

A second problem of eternal inflation is the initial condition
problem. It is shown that although inflation can be eternal to the
future, it can not be eternal to the past \cite{nopast}. There have
to be an initial condition for eternal inflation. The initial
condition of the universe may be given either at the quantum
creation of the universe \cite{Hartle:1983ai, Vilenkin:1984wp} or at
the start of the eternal inflation \cite{Cai:2007et}. It is not
clear whether the measure of eternal inflation should depend on the
initial conditions. Some authors believe that eternal inflation
should be independent of initial conditions \cite{vilenkin-measure},
while there are also calculations with results showing dependence on the
initial conditions.
\cite{Bousso:2006ev, Podolsky:2007vg}.

%Another problem is what kind of predictions can we make in eternal
%inflation. A lot of attempts have been made to provide an initial
%condition for the non-eternal stage of inflation from the
%probability in eternal inflation. But a lot still needs to be done
%in this way to make predictions and examine eternal inflation from
%experiments.

In this paper, we use the stochastic method \cite{s1,Gratton:2005bi}
to investigate the measure for the slow roll eternal inflation. This
method provides a possible solution to the problems listed above.
%, and make some predictions for the third problem.
We construct a local measure for the slow roll eternal inflation. In
the model with a scalar potential $\lambda\varphi^4$, it can be shown that the
measure for the low energy scale regime of eternal inflation is
independent of the initial condition. On the other hand, when the
energy scale of eternal inflation is higher than the scale where the
initial condition is proposed, the measure is initial condition
dependent. The de Sitter entropy arises naturally in this
situation.

As an application, this approach can be used to calculate the
probability for the inflaton to fluctuate from one local minimum to
another. The probability from this approach agrees with the tunneling
probability due to the Coleman-de Luccia instanton. Thus , this approach offers
a means to deal with the slow roll eternal inflation and the
tunneling eternal inflation in a single framework.

This paper is organized as follows. In Section 2, we review the
stochastic approach \cite{Gratton:2005bi} to eternal inflation. In
Section 3, we calculate the probability for the universe to have a
given Hubble radius, and discuss the physical implications for this
probability. We conclude in section 4.

\section{Gravity and a stochastic scalar field}
First, we review briefly the stochastic approach to
eternal inflation \cite{Gratton:2005bi}. The slow-roll condition can
be imposed self-consistently and one of the Friedmann equations takes the
usual form
\begin{equation}
  3H^2=V\ ,\label{friedmann}
\end{equation}
where we have set $8\pi G=1$. The result of quantum fluctuation of the
inflaton field can be mimicked by a Gaussian white noise
\begin{equation}
  3H\dot\varphi+V_\varphi=-H^{5/2}\eta(t)\ ,\label{varphieom}
\end{equation}
where $\eta(t)$ is Gaussian and normalized as
\begin{equation}
  <\eta(t)>=0\ ,\ \ \
  <\eta(t)\eta(t')>=\frac{9}{4\pi^2}\delta(t-t')\ .
\end{equation}
With such a normalization, the expectation value for a quantity
${\cal O}[\eta]$ is
\begin{equation}
  <{\cal O}[\eta]>=\int [d\eta]{\cal O}[\eta]\exp\left(-\frac{2}{9}\pi^2\int_0^\infty
  dt_1\eta^2(t_1)\right) \ .
\end{equation}
So one can recover the well-known result
\begin{equation}
<\delta_q\varphi^2>\simeq\frac{H^2}{4\pi^2}\ ,
\end{equation}
where $\delta_q\varphi$ is the quantum fluctuation during one Hubble
time and averaged in one Hubble volume.

For the potential $V=\lambda\varphi^4$, there exists an explicit
solution to the equations (\ref{friedmann}) and (\ref{varphieom}).
We define the Hubble length $R\equiv
1/H=\sqrt{\frac{3}{\lambda}}\frac{1}{\varphi^2}$, then the equations
(\ref{friedmann}) and (\ref{varphieom}) can be written as
\begin{equation}
  \dot R-\alpha R=\beta\eta(t)\ ,
\end{equation}
where $\alpha\equiv 8\sqrt{\lambda/3}$ and $\beta\equiv
2\sqrt[4]{\lambda/3}/3$. Given the initial condition $R=r_0$ when
$t=0$, the solution to the above equation is
\begin{equation}
  R(t)=r_0e^{\alpha t}+\beta e^{\alpha t}\int_0^t dt_1e^{-\alpha
  t_1}\eta(t_1) \ .
\end{equation}

\section{The probability density and its implications}

We now define and calculate the probability for eternal inflation
to enter a given region in the history space. When we consider a
spacially flat universe, using the slow-roll approximation, and
averaging the inflaton field over one Hubble volume, the history
space is parameterized by one single parameter. For simplicity, we
choose this parameter as the Hubble length $R=1/H$. Then the
probability $dP_{R_0}$ for eternal inflation to enter a region with
the Hubble length from $R_0$ to $R_0+dR_0$ can be written as
\begin{equation}
dP_{R_0}=P(R_0)dR_0 \ .
\end{equation}

The probability density $P(R_0)$ counts the number of times the
universe crosses the $R(t)=R_0$ surface during a infinite length of
time. So for a given function $\eta(t)$, it is proportional to an
integration of delta functions. Since $\eta(t)$ is stochastic, we
average over all possible $\eta(t)$ with the appropriate weight.
Then $P(R_0)$ takes the form
\begin{equation}
  P(R_0)\sim \int [d\eta]\exp\left({-\frac{2}{9}\pi^2\int_0^\infty dt_1\eta^2(t_1)}\right)\int_{t=0}^\infty dt\  \delta\left(R(t)-R_0\right)
  .\label{fi}
\end{equation}
Each time the universe across $R(t)=R_0$, the probability density
$P(R_0)$ picks up a contribution of delta function.

We pause to comment that in the above definition, $R_0$ can be replaced by
any other physical quantity if we are interested in computing the
probability distribution of this quantity.

It is in general not straightforward to calculate the functional
integration (\ref{fi}). While the calculation becomes relatively
easy when we consider the $\lambda\varphi^4$ theory. In this case,
we make use of the integration expression for delta function, and
approximate the continuous variable $t$ by a infinite number of
small time intervals $\Delta t$. In the last step we integrate out the
Gaussian integrals and take the $\Delta t\rightarrow 0$ limit. Then
the probability density takes the form
\begin{equation}
  P(R_0)\sim \int_0^\infty dt \sqrt{\frac{8\pi}{e^{2\alpha
  t}-1}}\exp\left(-8\pi^2r_0^2\frac{\left(e^{\alpha t}-\frac{R_0}{r_0}\right)^2}{e^{2\alpha
  t}-1}\right)\ . \label{pexpression}
\end{equation}
When $R_0\neq r_0$, the integration (\ref{pexpression}) is finite,
and the function $\frac{\left(e^{\alpha
t}-\frac{R_0}{r_0}\right)^2}{e^{2\alpha
  t}-1}$ on the expontential has two saddle points $e^{\alpha t}=\frac{R_0}{r_0}$ and $e^{\alpha t}=\frac{r_0}{R_0}$. We shall investigate
separately the $R_0>r_0$ and $R_0<r_0$ behavior of the integration
except the region where $R_0-r_0$ is much smaller than the Planck
length.

When $R_0>r_0$, let $e^{\alpha t}=\frac{R_0}{r_0}(1+x)$, then the
integration becomes
\begin{equation}\label{int1}
  P(R_0)\sim\int dx
  \frac{2\sqrt{2\pi}}{\alpha\left(1+x\right)\sqrt{(1+x)\left(\frac{R_0}{r_0}\right)^2-1}}\exp\left(-8\pi^2r_0^2\frac{x^2}{(1+x)^2-\left(\frac{r_0}{R_0}\right)^2}\right)
  \ .
\end{equation}
Since the integration is suppressed by a large exponential factor
$-8\pi^2r_0^2$,and $r_0$ need to be
larger than $1$ in the Planck units in order to neglect effects of quantum
gravity. So the integral is sharply peaked
at the saddle point. So this integration can be approximated by
\begin{equation}\label{int2}
  P(R_0)\sim\int dx
  \frac{2\sqrt{2\pi}}{\alpha\sqrt{\left(\frac{R_0}{r_0}\right)^2-1}}\exp\left(-8\pi^2r_0^2\frac{x^2}{1-\left(\frac{r_0}{R_0}\right)^2}\right)
  \ .
\end{equation}

It can be checked that the next to leading order correction (of the form $x^2$)
from (\ref{int1}) is suppressed by a factor $1/(8\pi^2r_0^2)$. So
(\ref{int2}) is a good approxiamtion to (\ref{int1}). The integral (\ref{int2}) can be worked out
to be
\begin{equation}\label{largeR}
  P(R_0)\sim\frac{1}{\alpha R_0}\ .
\end{equation}

The probability density (\ref{largeR}) is independent of the initial
condition $r_0$. This result is in
agreement with \cite{vilenkin-measure}. There are also some results
in which the probability distribution depends on
the initial condition \cite{Bousso:2006ev, Podolsky:2007vg}. However the
methods and models used there are different from ours.

Note that the $R_0>r_0$ region is allowed by the classical motion
without the random source $\eta(t)$. So it makes sense to compare
the result (\ref{largeR}) with the pure classical result. In the case
without the noise, the probability is
\begin{equation}
  P_{\rm cl}(R_0)\sim\int_{t=0}^{\infty} dt\ \delta(R(t)-R_0)\ .
\end{equation}
where $R(t)=r_0e^{\alpha t}$. The function $R(t)$ always increases
with $t$, and one obtain
\begin{equation}\label{largeRclassical}
  P_{\rm cl}(R_0)\sim\frac{1}{\partial_t R(t_0)}=\frac{1}{\alpha R_0}\ .
\end{equation}
This classical result is natural because $P_{\rm cl}(R_0)dR_0$ just
measures the proper time for the universe to stay between $R_0$ and
$R_0+dR_0$. Nevertheless one should not take this for granted for
other models.

The probability distribution with the random source (\ref{largeR}) is the
same as the classical probability density (\ref{largeRclassical}) in a
good approximation. So in this classically allowed region, the quantum
fluctuations do not change the result very much. This result is in
agreement with \cite{Gratton:2005bi}, in which the quantities such as
the e-folding number with quantum fluctuations are calculated and it
is shown that the quantum corrections are small.

On the other hand, when $R_0<r_0$, let $e^{\alpha
t}=\frac{r_0}{R_0}(1+x)$, then using a similar saddle point
approximation,
\begin{equation}\label{int3}
  P(R_0)\sim \int dx
  \frac{2\sqrt{2\pi}}{\alpha\sqrt{\left(\frac{r_0}{R_0}\right)^2-1}}\exp\left(-8\pi^2R_0^2\left(\left(\frac{r_0}{R_0}\right)^2-1+\frac{x^2}{1-\left(\frac{R_0}{r_0}\right)^2}\right)\right)\
  ,
\end{equation}
and (\ref{int3}) can be integrated out to give
\begin{equation}\label{smallR}
  P(R_0)\sim \frac{1}{\alpha r_0}
  e^{-8\pi^2(r_0^2-R_0^2)}\
  .
\end{equation}

\begin{figure}
\begin{center}
\includegraphics[width=5.5in]{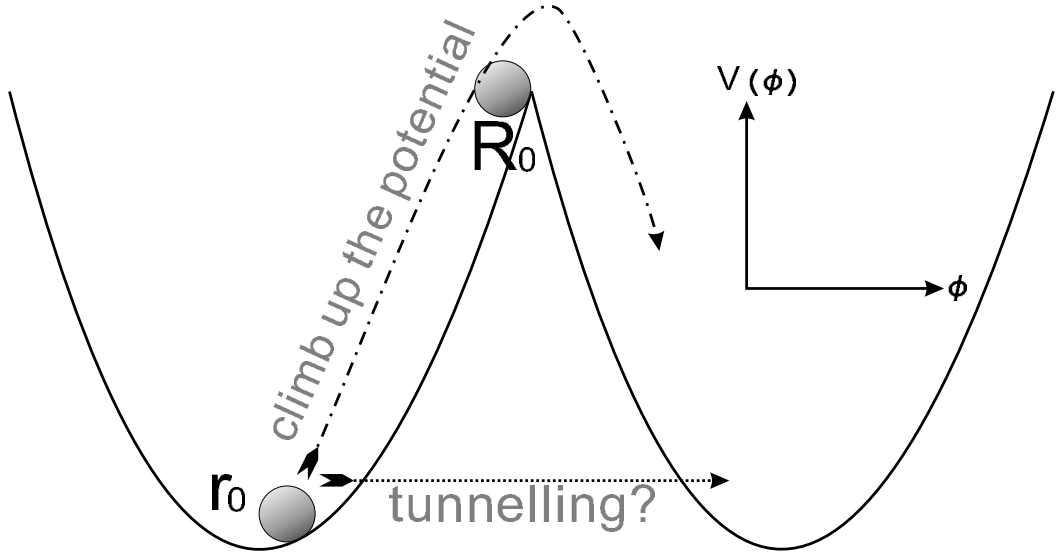}
\end{center}
\caption{Scalar field dynamics in the double well potential. Around
each minimum, the potential looks like
$\lambda(\varphi-\varphi_i)^4$ ($i=1,2$). The probability for a
stochastic scalar field to climb from one minimal to another agrees
with the quantum calculation using the Coleman de Luccia instanton.}
\label{fig1}
\end{figure}

This result also has interesting physical implications. Note that
$8\pi^2R^2$ is just the entropy of the de Sitter space with Hubble
radius $R$. So from the probability density (\ref{smallR}), we see
that the probability for the universe to fluctuate from a high de
Sitter entropy state to a low entropy state is suppressed by the
exponential of the minus entropy difference. This result is in
agreement with the generalized second law of thermodynamics and the
calculation made in \cite{Bousso:2006ev}. And as in
\cite{Arkani-Hamed:2007ky}, it provides another operational meaning
to the de Sitter entropy.

As a special case, let us consider the probability for the universe
to tunnel from one $\lambda\varphi^4$ like minimum to another (see
Fig. \ref{fig1}). Initially, the universe stays near one minimum of
the potential. If $r_0\gg R_0$, the probability for the inflaton to
randomly climb up the potential and get to the other minimum is
suppressed by the factor $\exp (-8\pi^2 r_0^2)$. This agrees with the
calculation using the Coleman de Luccia instanion
\cite{Coleman:1980aw}.

\section{Conclusion}

In this paper, we used a stochastic source to simulate the quantum
fluctuation of the inflaton. We defined the probability for
the universe to be at any given Hubble radius. It is shown in a concrete
model that the probability can be calculated when the
difference between $r_0$ and $R_0$ is larger than the Planck length.

When $R_0>r_0$, the probability is dominated by the classical
probability without the random source, and the quantum correction is
suppressed by the factor $1/(8\pi^2r_0^2)$. While in the classical
forbidden region $r_0>R_0$, the probability is suppressed by the
exponential of the minus entropy difference.

Our definition of the measure and the calculation of the
probability offers a possible solution to the measure problem in
inflation, and may lead to some insight to the physical meaning for the entropy
of the de Sitter space. Although explicit calculations are
performed in a single field inflation model with a $\lambda\varphi^4$
potential, the results have clear physical meaning, thus appear
quite general, it remains an open problem whether the
stochastic multi-field model with more general potentials share the
nice features demonstrated in this paper.

\section*{Acknowledgments}
This work was supported by grants of NSFC. We thank Yi-Fu Cai,
Chao-Jun Feng, Wei Song and Yushu Song for discussions.

\end{document}